# Stability Mechanisms of Unconventional Stoichiometric Crystals Exampled by Two-Dimensional Na$_2$Cl on Graphene under Ambient Conditions

Liuhua Mu[1,2], Xuchang Su[3], Haiping Fang[4,5,6]*, Lei Zhang[3]†

[1]*Wenzhou Institute, University of Chinese Academy of Sciences, Wenzhou, Zhejiang 325001, China*
[2]*School of Physical Science, University of Chinese Academy of Sciences, Beijing 100049, China*
[3]*MOE Key Laboratory for Nonequilibrium Synthesis and Modulation of Condensed Matter, School of Physics, Xi'an Jiaotong University, Xi'an 710049, China*
[4]*School of Physics, East China University of Science and Technology, Shanghai 200237, China*
[5]*Shanghai Institute of Applied Physics, Chinese Academy of Sciences, Shanghai 201800, China*
[6]*School of Physics, Zhejiang University, Hangzhou 310027, China*

Compounds harboring active valence electrons, such as unconventional stoichiometric compounds of main group elements including sodium, chlorine, and carbon, have conventionally been perceived as unstable under ambient conditions, requiring extreme conditions including extra-high pressure environments for stability. Recent discoveries challenge this notion, showcasing the ambient stability of two-dimensional Na$_2$Cl and other unconventional stoichiometric compounds on reduced graphene oxide (rGO) membranes. Focusing on the Na$_2$Cl crystal as a case study, we reveal a mechanism wherein electron delocalization on the aromatic rings of graphene effectively mitigates the reactivity of Na$_2$Cl, notably countering oxygen-induced oxidation—a phenomenon termed the Surface Delocalization-Induced Electron Trap (SDIET) mechanism. Theoretical calculations also show a substantial activation energy barrier emerges, impeding oxygen infiltration into and reaction with Na$_2$Cl. The remarkable stability was further demonstrated by the experiment that Na$_2$Cl crystals on rGO membranes remain almost intact even after prolonged exposure to a pure oxygen atmosphere for 9 days. The discovered SDIET mechanism presents a significant leap in stabilizing chemically active substances harboring active valence electrons under ambient conditions. Its implications transcend unconventional stoichiometric compounds, encompassing main group and transition element compounds, potentially influencing various scientific disciplines.

Compounds of main group elements, whether adhering to conventional or unconventional stoichiometries, constitute the vast array of substances in the natural world. Unconventional stoichiometric compounds of main group elements exhibit distinctive electronic properties [1-10], with partial active valence electrons that do not obey the 8-valence-electron rule (i.e., octet rule). For example, in unconventional stoichiometric Na$_2$Cl, two Na atoms contribute two *s*-orbital valence electrons, whereas ideally only one should combine with one Cl atom, leaving the other as an active valence electron to potentially cause structural instability. Correspondingly, Na$_2$Cl and Na$_3$Cl crystals, termed as "violating the octet rule" [6], were believed to only be synthesized through extra-high pressure experiments at above 100 Gpa and 77 GPa, respectively [2]. Therefore, unconventional stoichiometric compounds of the main group elements had traditionally been viewed as structurally unstable under ambient conditions due to their unbalanced stoichiometries and the elevated chemical reactivity from their valence electrons in *s*-orbitals or *p*-orbitals [11]. Consequently, synthesis of those compounds in the past usually through extra-high pressure experiments in a laser-heated diamond anvil cell (e.g., Na$_2$Cl, Na$_3$Cl, CaO$_3$, Na$_2$HeO) [2,7,8], and under ambient conditions they cannot maintain stability. Clearly, the abnormal valence electrons that violate the octet rule in the compounds give rise to unique structural and physical properties, playing a pivotal role in advancing both scientific understanding and practical applications. Nevertheless, under ambient conditions, the practical utility of these compounds faces significant obstacles, due to the challenge of preserving their structural stability.

One would not expect unconventional stoichiometric compounds of main group elements, such as Na$_2$Cl, Na$_3$Cl, and CaCl crystals, to exist naturally or be synthesized under ambient conditions. Recently, the direct observations of two-dimensional (2D) Na$_2$Cl, Na$_3$Cl, Li$_2$Cl, K$_2$Cl, and CaCl crystals [1,5,9,10] under ambient conditions on reduced graphene oxide (rGO) membranes have challenged these expectations. These crystals form at the liquid-solid interfaces in unsaturated salt solutions and remarkably retained stability even after being removed from the salt solutions and drying. Notably, the unbalanced stoichiometries in these compounds can result in exotic physical and structural properties. For example, unexpected ferromagnetism is present in both 2D Na$_2$Cl [12] and CaCl [5] crystals, and periodic atom vacancies exist in the 2D $Na_2Cl$ crystals. Moreover, the new method to get a high content of 2D $Na_2Cl$ crystals synthesized in rGO membranes [13] facilitates the practical applications of these exotic properties. Very recently, a piezoelectric sensor with atomic-scale sensitivity, superior to all current state-of-the-art pressure sensors and showcasing an extremely low detection limit of only 25 nanonewtons, has been fabricated, benefiting from the periodic atom vacancies in the high content 2D $Na_2Cl$ crystals with a very large polarization [14].

However, we still suffer from the apparent paradox between the stable existence of compounds (e.g., $Na_2Cl$) under ambient conditions and the simultaneous presence of chemically active valence electrons arising from unconventional stoichiometries.

In this study, taking the $Na_2Cl$ crystal as an example, we reveal a mechanism wherein electron delocalization on the aromatic rings of graphene effectively mitigates the reactivity of $Na_2Cl$. This phenomenon occurs through the confinement of active valence electrons, which are triggered by the strong ion–π interactions between the $Na_2Cl$ crystal and aromatic rings, notably countering oxygen-induced oxidation. We term this mechanism the Surface Delocalization-Induced Electron Trap (SDIET). To assess the degree of structural stability and oxidation resistance of $Na_2Cl$ crystals, we conducted experiments and found that $Na_2Cl$ crystals remained structurally intact with negligible variation for at least 9 days in an oxygen atmosphere. We note that the SDIET mechanism is universal, it holds the potential to achieve the stabilization of other chemically active compounds that were previously considered unstable under ambient conditions, in addition to unconventional stoichiometric compounds on graphene. Such compounds, featuring active valence electrons, could exhibit more exotic physical properties that extend far beyond our current understanding.

To explore the stabilization mechanism of the $Na_2Cl$ crystal on graphene (denoted as $Na_2Cl$+graphene, Fig. 1a), we conducted investigations into the electron density near the Fermi level (i.e., the distribution of valence electrons) by using density functional theory (DFT) calculations. DFT calculations were performed by using the VASP [15-17] and CP2K [18,19] programs with Perdew-Burke-Ernzerhof (PBE) exchange–correlation functional [20]. The reliability and efficiency of these programs in analyzing electron density have been confirmed in previous studies [21-28]. The more details of the calculations and the related parameters are described in PS1 of Supplemental Material. Valence electrons, which significantly influence the chemical reactivity of materials [29], are typically localized near the Fermi level. The distribution of electron density near the Fermi energy level from $E_F−0.4$ eV to $E_F$ in Fig. 1 shows the effective delocalization of valence electrons on the $Na_2Cl$ crystal into the extensive π-conjugated system of aromatic rings present on the graphene, indicating the strong ion–π interactions between the $Na_2Cl$ crystal and the aromatic rings. The distributions of electron density in other energy ranges (with a narrow range from $E_F−0.1$ eV to $E_F$, and a broader range from $E_F−1.1$ eV to $E_F$) below the Fermi level also show similar effective delocalization of the valence electrons on the $Na_2Cl$ crystal into the graphene (Fig. S1).

The effective delocalization of the valence electrons indicates that they are confined between the $Na_2Cl$ crystal and graphene, and thus the delocalization is favorable for the structural stability of the $Na_2Cl$ crystal, although they violate the octet rule induced by unbalanced stoichiometries (see more analyses about valence electron properties in PS2 of Supplemental Material and Fig. S2). Specifically, this efficient dispersal of valence electrons over a greater number of atomic centers diminishes the electron density concentration on any individual atom. This SDIET mechanism, in turn, reduces the chemical reactivity, ultimately leading to excellent structural stability of the $Na_2Cl$ crystal. Moreover, it appears that graphene plays a role in electronically stabilizing $Na_2Cl$+graphene similar to how N-heterocyclic carbenes (NHCs) stabilize radicals, as the structure of NHCs facilitates the delocalization of valence electrons within the molecule [30,31].

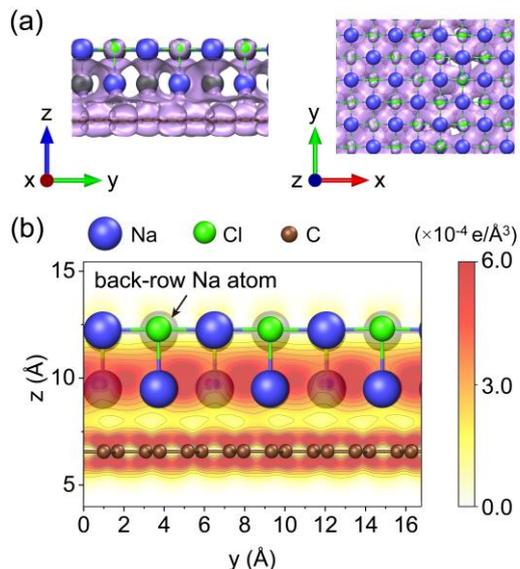

FIG. 1. Electron density analyses. (a) Supercell structure and electron density near the Fermi level (from $E_F−0.4$ eV to $E_F$) for the $Na_2Cl$ crystal on graphene ($Na_2Cl$+graphene). The Fermi level is the highest occupied level, denoted as $E_F$. The electron density is plotted for the iso-value of $2.3×10^{-4}$ e/Å$^3$. The upper and lower panels correspond to the side and top views, respectively. In the side view, back-row sodium atoms are represented as grey spheres for clarity. (b) Contour plots of the electron density near the fermi energy level (from $E_F−0.4$ eV to $E_F$), averaged in the y-z plane of $Na_2Cl$+graphene.

In order to evaluate the structural stability and oxidation resistance of the 2D Na$_2$Cl crystal, we firstly performed geometry optimizations of an oxygen molecule adsorbed onto the 2D Na$_2$Cl crystal sheet (referred to as O$_2$@(Na$_2$Cl+graphene)), with an oxygen concentration of approximately 0.3 O$_2$ molecules nm$^{-2}$. Within Na$_2$Cl+graphene, all four Na–Cl bonds exhibit uniform characteristics for each Na ion situated on the Na$_2$Cl crystal surface, with an average bond length of 2.8 Å (Fig. 2a). After the adsorption of an oxygen molecule, the average bond length experiences a slight increase from 2.8 Å in Na$_2$Cl+graphene to 2.9 Å in O$_2$@(Na$_2$Cl+graphene). This minor bond disproportionation in the local structure of the adsorbed Na site is attributed to the oxygen molecule exerting a slight influence on its four nearest Cl neighbors, directing them toward the graphene substrate. Consequently, the thickness of the Na$_2$Cl crystal in O$_2$@(Na$_2$Cl+graphene) reaches 3.6 Å, marginally larger than the 2.9 Å thickness of the Na$_2$Cl crystal in Na$_2$Cl+graphene. Collectively, these findings suggest that the adsorption of oxygen molecules induces limited structural distortions within the Na$_2$Cl crystal.

Subsequently, we evaluated the energetics of the oxidation process involving the 2D Na$_2$Cl crystal on graphene. Fig. 2b illustrates the relative energy profiles associated with the progression towards the formation of Na oxides. This process encompasses the chemical adsorption of an oxygen molecule ($O_2^{ads}$) onto the surface Na ion (stage 1, adsorption) and the infiltration of the oxygen molecule into the lattice to form an oxide (stage 2, oxide formation). Detailed configurations of these steps are provided in Fig. S3. Upon the adsorption of an oxygen molecule onto Na$_2$Cl+graphene, $O_2^{ads}$ is positioned 3.0 Å above the Na$_2$Cl crystal surface with an adsorption energy of −2.0 eV (Fig. 2b). Most importantly, $O_2^{ads}$ does not instantaneously trigger oxide formation at the surface due to the substantial energies required, amounting to 1.5 eV (Fig. 2b). This observation strongly implies that the oxidation process of the 2D Na$_2$Cl crystal on graphene should be significantly suppressed and attenuated in an oxygen-enriched atmosphere.

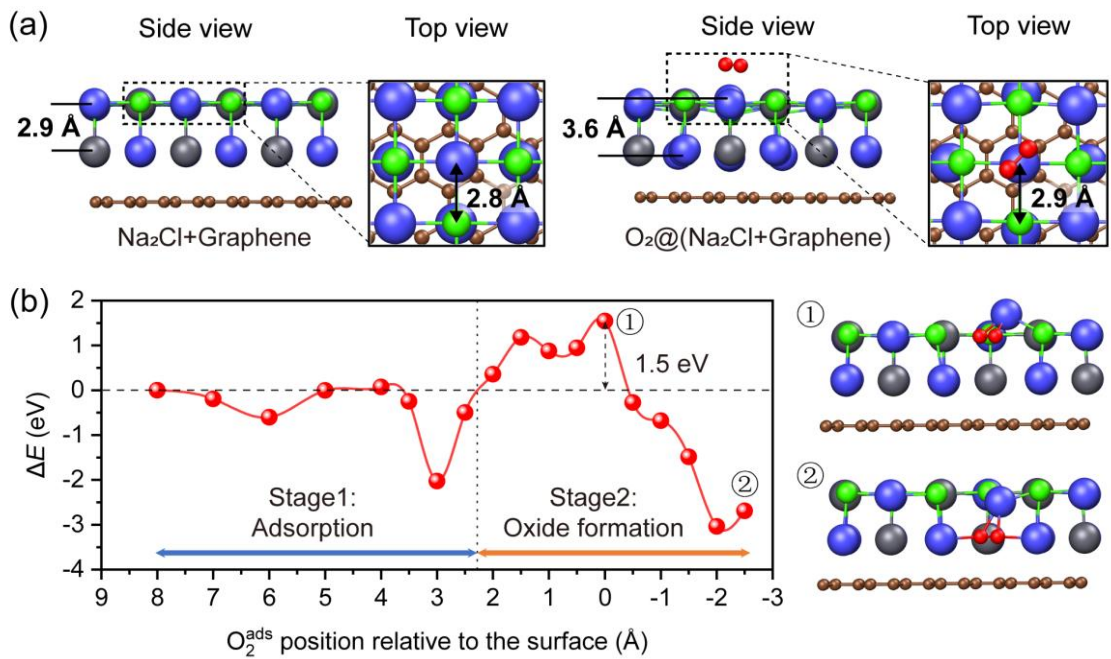

FIG. 2. DFT calculations on the mechanism of long-lasting oxidation resistance of the 2D Na$_2$Cl crystal on graphene (Na$_2$Cl+graphene). (a) Optimized structures of Na$_2$Cl+graphene with and without O$_2$ adsorption (O$_2$@(Na$_2$Cl+graphene)). (b) Relative energy profile of $O_2^{ads}$ adsorbed onto Na$_2$Cl+graphene. The position of $O_2^{ads}$ is depicted relative to the Na$_2$Cl surface plane. The plots in the right panel illustrate the two configurations involved in the oxidation process, identified as ① and ②, in alignment with the left panel. Spheres in red, blue, green, and brown represent oxygen, sodium, chlorine, and carbon, respectively. In side views, back-row sodium atoms are represented as grey spheres for clarity.

To gain further insight into the stability of Na$_2$Cl+graphene, we conducted DFT computations of the ionization potential (IP). The IP represents the energy required for removing an electron from a molecule, as schematically illustrated in the inset of Fig. 3, and is a crucial parameter for defining the chemical stability of a compound [32]. In this context, the calculated IP value was derived from the electron residing in the Na *s*-orbitals of Na$_2$Cl+graphene (see PS1 in Supplemental Material for more details). For comparison, we also calculated the IP value for the Na *s*-orbitals of Na metal (see geometrical structure in Fig. S4). Fig. 3 shows that the IP value for Na$_2$Cl+graphene is positive (1.3 eV), signifying that the extraction of valence electrons from Na$_2$Cl+graphene is an endothermic and unfavorable process. This observation strongly suggests a long-lasting oxidation resistance of Na in an oxygen atmosphere. In contrast, the IP value for Na metal is negative (−0.4 eV), as shown in Fig. 3, implying that the extraction of valence electrons from Na metal is an endothermic process, which is consistent with the well-documented high chemical reactivity of Na metal [33]. Moreover, it has been reported that compounds exhibiting strong electronic delocalization tend to possess elevated IP values [32,34]. Thus, it is noteworthy that the relatively high IP value of Na$_2$Cl+graphene can be attributed to its effective delocalization of electrons across the extensive π-conjugated system of graphene (Fig. 2).

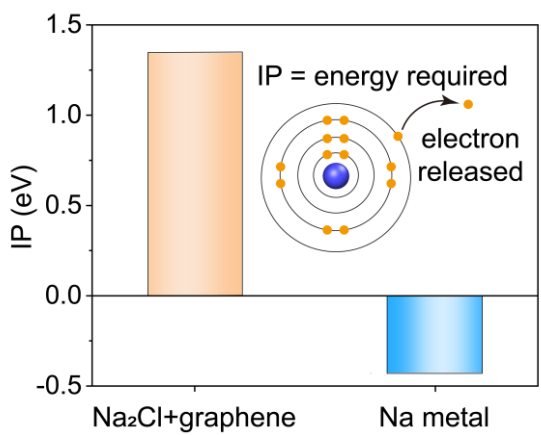

FIG. 3. Calculated Ionization potentials (IP) of Na$_2$Cl+graphene and Na metal. The inset shows a schematic representation of the ionization potential definition.

In order to assess the degree of structural stability and oxidation resistance of Na$_2$Cl crystals, as explored by the above theoretical analyses, especially the SDIET mechanism (shown schematically in Fig. 4a), we embarked on experimental investigations.

Firstly, we fabricated rGO membranes containing 2D Na$_2$Cl crystals following our previous methodologies [1]. Freestanding rGO membranes [1,13,35-37] were prepared through the drop-casting and thermal reduction methods from graphene oxide suspension (see PS1 in Supplemental Material for details). Subsequently, these membranes were immersed in 3.8 mol l$^{-1}$ (M) NaCl solutions at room temperature for 48 hours. The resulting wet rGO membranes with salt solution were subjected to X-ray diffraction (XRD) analysis. As illustrated in Fig. 4b, numerous Bragg peaks were observed at distinct diffraction angles (2θ). The Bragg peaks at ~19° and 25° correspond to the rGO membranes, while the Bragg peak at ~32° corresponds to the (200) surface of the 2D Na$_2$Cl crystals, consistent with our previous XRD analyses [1]. This result unequivocally confirms the formation of the 2D Na$_2$Cl crystals on the rGO membranes under ambient conditions.

Subsequently, we scrutinized the oxidation resistance of the Na$_2$Cl crystals on rGO membranes in an oxygen atmosphere using X-ray photoelectron spectroscopy (XPS) and electron spin resonance (ESR) spectroscopy. The rGO membranes containing 2D Na$_2$Cl crystals, here denoted as Na$_2$Cl@rGO, were subjected to drying under vacuum at 70 ºC for six hours and subsequently sealed within an oxygen atmosphere for 0, 9, and 30 days. These Na$_2$Cl@rGO were then analyzed using XPS to assess their oxidation resistance. The temporal evolution of their chemical compositions is presented in Fig. 4c (additional details provided in PS3 of Supplemental Material and Fig. S5). Remarkably, the Na:Cl ratio of 2.0 remained essentially unchanged even after 9 days of exposure to oxygen atmosphere (Fig. 4c). Further XPS analyses of atomic oxygen concentrations confirmed the ingress of oxygen into the interior of the rGO membranes (Fig. S6). It is worth noting that a Na:Cl ratio of 2.0 demonstrates the formation of Na$_2$Cl crystals on graphene, consistent with our prior observations [1,13]. Beyond 30 days of exposure, the Na:Cl ratio increased to 3.0 (Fig. 4c), signifying a mild oxidation of Na$_2$Cl@rGO. Consequently, the constant Na:Cl ratio (i.e., chemical compositions) over extended periods, up to 9 days in an oxygen atmosphere, underscores the long-lasting oxidation resistance and sustainable, long-term oxidative stability of Na$_2$Cl@rGO under such conditions.

The enduring oxidation resistance of Na$_2$Cl@rGO was further validated through ESR spectroscopy experiments, which provide insights into the unusual electronic properties of Na$_2$Cl@rGO. Fig. 4d illustrates the evolution of ESR signals for Na$_2$Cl@rGO as a function of exposure time in an oxygen atmosphere. Na$_2$Cl@rGO, exhibiting a substantial number of unpaired valence electrons, was characterized by pronounced ESR responses (Fig. 4d). These ESR responses were not observed in the control system (rGO membranes, see Fig. S7), which indicates that the ESR responses of the unpaired valence electrons mainly come from Na$_2$Cl crystals. Importantly, a relatively stable intensity of these signals was observed during the initial 9 days of exposure to the oxygen atmosphere (Fig. 4d), suggesting a limited effect of oxygen on the electronic properties of Na$_2$Cl@rGO. This remarkable stability in electronic properties holds promise for the design of novel transistor devices.

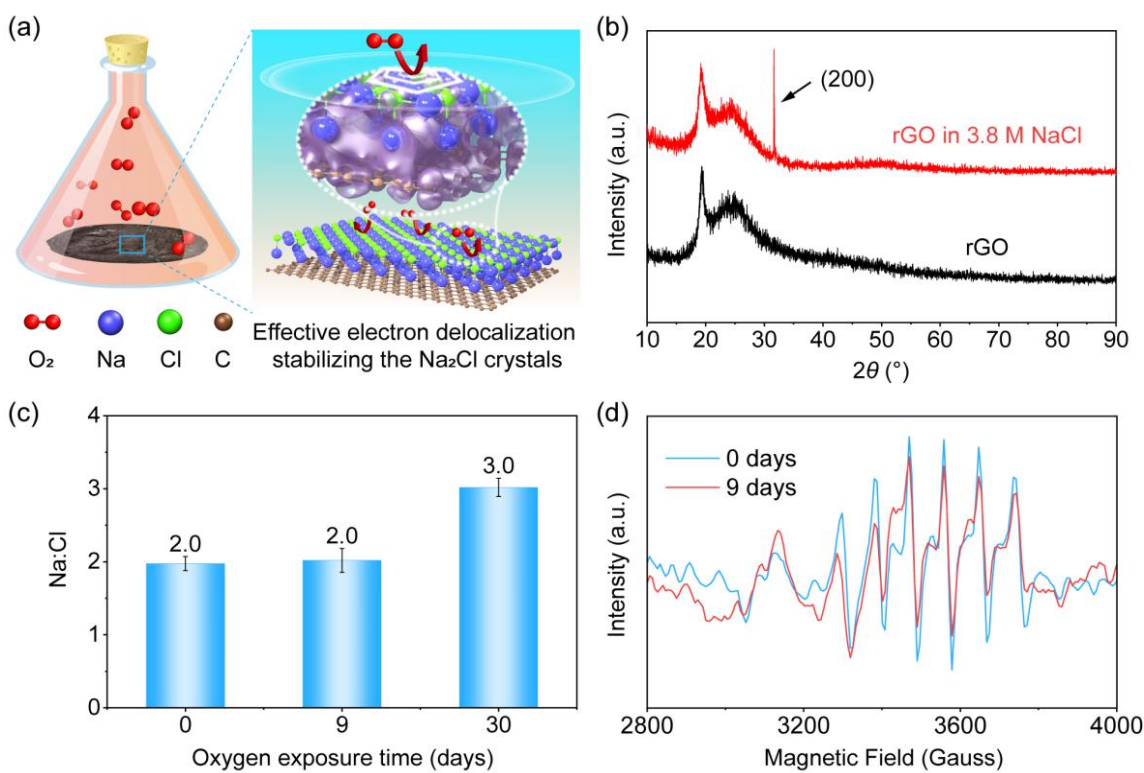

FIG. 4. Long-lasting oxidation resistance of Na$_2$Cl@rGO in an oxygen atmosphere. (a) Illustration of the experimental setup and fundamental mechanisms. Effective electron delocalization plays a pivotal role in safeguarding the surface against attack by oxygen. Inset: the red arrows represent the oxidation resistance exhibited by Na$_2$Cl@rGO. The purple iso-surface illustrates the electron density of the highest occupied molecular orbitals of the Na$_2$Cl crystal on the graphene substrate. (b) XRD patterns of the rGO membranes immersed in the 3.8 M NaCl solution together with the XRD patterns of the control system (rGO membranes). (c) Atomic ratio of Na to Cl in Na$_2$Cl@rGO as a function of oxygen exposure time. (d) ESR spectra of Na$_2$Cl@rGO following exposure to oxygen atmosphere for 0 and 9 days.

In this study, our theoretical investigations have identified a mechanism that effectively curtails the chemical reactivity of active valence electrons, resulting in heightened stability of Na$_2$Cl crystals on graphene even when exposed to an oxygen atmosphere for 9 days. This mechanism, termed SDIET, is revealed by the effective delocalization of electrons across the extensive π-conjugated framework of graphene, which facilitates the confinement of active valence electrons, thereby countering oxygen-induced oxidation and preserving the structural integrity of Na$_2$Cl crystals on graphene.

The unconventional stoichiometric compounds, housing active valence electrons, are induced by the imbalance in the strength of the interaction between the cations and anions with the substrate, here are mainly cation-π and anion-π interactions. The prevalence of cation-π and anion-π interactions, not only with sodium and chlorine ions but also with a variety of other ions (e.g., Li, K, Mg, Ca, F, Br, and I) on graphitic surfaces [37-41], widens the scope of compounds with unconventional stoichiometry. This diversity underscores the potential of the SDIET mechanism to extend its stabilizing influence to various chemically active compounds formed by main group elements.

Moreover, unconventional stoichiometric compounds are not limited to graphitic surfaces with π electrons to form ion-π interactions. Other types of surfaces, exhibiting different but distinct non-covalent interactions with cations and anions, may also induce unconventional stoichiometric compounds. Very recently, Na$_2$Cl and Na$_3$Cl crystals have also been observed on the iron surface due to metal-ion interactions under ambient conditions [42]. Clearly, the SDIET mechanism can broaden its applicability in those unconventional stoichiometric compounds with active valence electrons.

Now we come to the transition elements with more intricate valence electron structures. Through the SDIET mechanism, active valence electrons may also be confined, and there are interactions between those confined active valence electrons with inner electrons. The corresponding compounds could manifest a broader range of updated characteristics while maintaining stable structures, potentially surpassing our current understanding.

In conclusion, the discovered SDIET mechanism provides a comprehensive framework for confining valence electrons to achieve the stabilization of chemically active substances encompassing main group and transition element compounds. Triggered by ion-π interactions or other non-covalent interactions under ambient conditions, this understanding offers prospects for understanding unique material properties and incites widespread interest, as well as leads to innovative applications in various technological domains, empowering the creation of novel electronic, magnetic, optical, and mechanical devices at the atomic scale.

We thank Guosheng Shi and Liang Chen for their constructive suggestions. This work is supported by the National Natural Science Foundation of China (No. 11974366), the Postdoctoral Fellowship Program of CPSF under Grant Number GZC20232610, the Fundamental Research Funds for the Central Universities, and Young Talent Support Plan of Xi'an Jiaotong University. We thank the Shanghai Supercomputer Center for computing time. We also thank the Instrument Analysis Center of Xi'an Jiaotong University for XPS analyses.

---

†Corresponding author: zhangleio@mail.xjtu.edu.cn
*Corresponding author: fanghaiping@sinap.ac.cn